\documentclass[preprint,12pt]{elsarticle}

\journal{Journal of Quantitative Spectroscopy and Radiative Transfer}
\usepackage{amssymb}
\usepackage{graphicx}
\usepackage{epsfig}
\usepackage{color}
\usepackage{rotating}
\usepackage{float}
\usepackage{epstopdf}
\usepackage{array}
\usepackage{threeparttable}

\usepackage{caption}

\newcommand{\etal}{\textit{et al.}}

\newcommand{\subtwo}[2] {${\rm{#1_{#2}}}$}
\newcommand{\supertwo}[2] {${\rm{#1^{#2}}}$}
\newcommand{\CC}{C\sub{2}}
\newcommand{\super}[1] {\supertwo{}{#1}}
\newcommand{\sub}[1] {\subtwo{}{#1}}
\newcommand{\wavenumbers}{cm\super{-1}}
\newcommand{\lowerstate}{a\super{3}$\Pi$\sub{u}}
\newcommand{\upperstate}{d\super{3}$\Pi$\sub{g}}
\newcommand{\tPI}{\super{3}$\Pi$}
\newcommand{\Bv}{B\sub{\emph{v}}}
\newcommand{\Gv}{G(\emph{v})}

\newcommand{\primed}{\super{\prime}}
\newcommand{\bSigma}{b\super{3}$\Sigma\rm{^-_g}$}
\newcommand{\OnePi}{1\super{5}$\Pi\rm{_g}$}

\newcommand{\Avv}{Einstein \textit{A}$_{v^{\prime}}$$_{v^{\prime\prime}}$}
\newcommand{\fvv}{$f_{v^{\prime}}$$_{v^{\prime\prime}}$}

\begin{document}

\begin{frontmatter}

\title{Line strengths and updated molecular constants for the \CC\ Swan system}

\author{James S. A. Brooke\corref{cor1}}
\address{Department of Chemistry, University of York, York, YO10 5DD,
UK; jsabrooke@gmail.com; +44 1904 434525.}
\cortext[cor1]{Corresponding Author}

\author{Peter F. Bernath}
\address{Department of Chemistry \& Biochemistry, Old Dominion University, 4541 Hampton Boulevard, Norfolk, VA, 23529-0126, USA; and Department of Chemistry, University of York, Heslington, York, YO10 5DD,
UK.}

\author{Timothy W. Schmidt and George B. Bacskay}
\address{School of Chemistry, Building F11, The University of Sydney, New South Wales, 2006, Australia.}

\begin{abstract}
New rotational line strengths for the \CC\ Swan system (\upperstate-\lowerstate) have been calculated for vibrational bands with \textit{v}\primed=0-10 and \textit{v}\primed\primed=0-9, and \textit{J} values up to \textit{J}=34-96, based on previous observations in 30 vibrational bands. Line positions from several sources were combined with the results from recent deperturbation studies of the \textit{v}\primed=4 and \textit{v}\primed=6 states, and a weighted global least squares fit was performed. We report the updated molecular constants. The line strengths are based on a recent ab initio calculation of the transition dipole moment function. A line list has been made available, including observed and calculated line positions, Einstein \textit{A} coefficients and oscillator strengths (\textit{f}-values). The line list will be useful for astronomers and combustion scientists who utilize \CC\ Swan spectra. Einstein \textit{A} coefficients and \textit{f}-values were also calculated for the vibrational bands of the Swan system.

\end{abstract}


\end{frontmatter}

\section{Introduction}
\CC\ is an important molecule in the fields of astronomy, combustion science and materials science. It has often been observed in comets~\citep{1968Mayer-a, 1976Jackson-a, 1983Lambert-a, 1983Johnson-a, 1997Sorkhabi-a, 2003Kaiser-a} and in other astronomical environments such as interstellar clouds~\cite{1977Souza-a,1978Chaffee-a,1981Green-a,1982Hobbs-a,1989Federman-a,2010Kazmierczak-a,2012Casu-a}, late-type stars~\citep{1970Vardya-a,1971Querci-a,2000Klochkova-a,2012Hema-a} and the Sun~\citep{1973Grevesse-a,1982Brault-a}. Its reactions are believed to be involved in the formation of hydrocarbons and other organic compounds in interstellar clouds~\citep{2002Kaiser-a}. \CC\ has also been found in flames~\citep{1957Gaydon-Book-a,1965Bleekrode-a,1977Baronavski-a} and from the irradiation of soot~\citep{2010Goulay-a}, and can be formed in carbon plasmas and used to make carbon nanostructures~\citep{2011Nemes-Book-a}.

The most prominent electronic system in the visible region is the Swan system, which involves the electronic transition \upperstate-\lowerstate, with the (0,0) band near 19400 \wavenumbers. The \lowerstate\ state was originally believed to be the ground state~\citep{1969Herzberg-a} as it was observed to be easily excited, which is because it lies only 1536 \wavenumbers\ above the actual X\super{1}$\Sigma^+_g$ ground state.

The Swan system has been investigated extensively. Early vibrational band intensity analyses include those of King~\citep{1948King-a}, Phillips~\citep{1957Phillips-a} and Hagan~\citep{1963Hagan-Report-a}. In 1965, Mentall and Nicholls~\citep{1965Mentall-a} reanalysed the data of three previous works to provide an updated list of absolute band strengths, oscillator strengths and Einstein \textit{A} values for most vibrational bands up to {\em{v\primed}}=4. A full review of previous work was given in 1967 by Tyte \emph{et al.}~\citep{1967Tyte-Report-a}. In 1968, Phillips and Davis~\citep{1968Phillips-Book-a} combined earlier published data with their most recent rotational analysis. They calculated spectroscopic constants for the Swan system, and published a full rotational line list including relative intensities. Danylewych and Nicholls published a list of absolute band strengths, oscillator strengths and Einstein \textit{A} values covering most vibrational bands of up to {\em{v\primed}}=9, with $\Delta${\em{v}} $\leq$ 4~\citep{1974Danylewych-a}. The properties of \CC\ were extensively reviewed by Huber and Herzberg in 1979~\citep{1979Huber-Book-a}.

As new experimental techniques have become available, new studies of the lower vibrational bands have been conducted at high resolution. These include the work of Amiot~\citep{1983Amiot-a}, Curtis and Sarre~\citep{1985Curtis-a} and Suzuki \emph{et al.}~\citep{1985Suzuki-a}, who investigated the (0-0), (0-1) and (1-0) bands, respectively, using laser excitation techniques. Dhumwad \emph{et al.}~\citep{1981Dhumwad-a} observed the Swan system using a quartz discharge tube with tungsten electrodes for the excitation of CO. In 1994, Prasad and Bernath~\citep{1994Prasad-a} analysed nine low vibrational bands ({\em{v\primed}} $\leq$ 3 and {\em{v\primed\primed}} $\leq$ 4) of the Swan system of jet-cooled \CC\ (for low-J), and of \CC\ produced in a composite wall hollow cathode (for \textit{J} up to 25-46) with a Fourier Transform Spectrometer (FTS). The previous observations of Amiot \etal\ and Prasad and Bernath on the (0,0) band were improved upon by Lloyd and Ewart in 1999~\citep{1999Lloyd-a}, using degenerate four-wave mixing spectroscopy. These and other investigations have improved the accuracy of the line assignments originally published by Phillips and Davis in 1968~\citep{1968Phillips-Book-a}.

The higher vibrational bands had not been analyzed with modern high resolution instrumentation until 2002, when Tanabashi and Amano~\citep{2002Tanabashi-a} observed the Swan system by a direct absorption technique using a tunable dye laser. They measured three bands, assigned as (7,9), (6,8) and (5,7). They found that their line positions did not agree with those reported by Phillips and Davis~\citep{1968Phillips-Book-a}, which was the most recent rotational analysis of the high vibrational bands. These discrepancies led to the reanalysis of the entire Swan system with a high resolution FTS~\citep{2007Tanabashi-a}. The assigned line positions from this new comprehensive analysis agreed with their previous one for the (7,9), (6,8) and (5,7) bands. Their line positions for bands involving the higher vibrational levels differed significantly from those of Phillips and Davis~\citep{1968Phillips-Book-a}.

The old rotational line list reported by Phillips and Davis in 1968~\citep{1968Phillips-Book-a} has recently been used in deriving the carbon abundance and \super{12}C/\super{13}C ratio in R Coronae Borealis and hydrogen-deficient carbon stars~\citep{2012Hema-a}, and in comets~\citep{2012Rousselot-a}. Line positions and intensities must be known to derive \CC\ abundance from observed spectra, and it would be beneficial to have a new rotational line list, based on recent measurements and calculations. The purpose of this work is to use the data mainly of Tanabashi \etal\ \citep{2007Tanabashi-a} to calculate theoretical line intensities, and publish an extensive line list. Similar work has recently been carried out for the E$^2\Pi$-X$^2\Sigma^+$ transitions of CaH by Li \etal\ \citep{2012-Li-a}

Tanabashi \emph{et al.} assigned around 5700 observed rotational lines, for 34 vibrational bands belonging to the $\Delta$\textit{v} = -3 to +2 sequences. Transitions up to between \textit{J}=30 and 80 were assigned. Perturbations were found in the \upperstate\ state for \emph{v} = 0, 1, 2, 4, 6, 8, 9 and 10, and for \emph{v} = 4, 6 and 9 they affected almost all of the observed lines. They calculated molecular constants (Tables 3 and 4 in ref~\citep{2007Tanabashi-a}) for both electronic states.

Deperturbation studies of the \upperstate\ \textit{v}\primed=4~\citep{2010Bornhauser-a} and \textit{v}\primed=6~\citep{2011Bornhauser-a} states were performed by Bornhauser \textit{et al.} in 2010 and 2011, respectively, using double-resonant four-wave mixing spectroscopy. This enabled them to assign lines unambiguously, and calculate perturbation constants (Table \ref{TABperts}) and molecular constants of the interacting \bSigma\ (\textit{v}=16 and \textit{v}=19) and \OnePi\ states. They also gave a list of the few transitions that were assigned incorrectly by Tanabashi \etal.

\section{Recalculation of Molecular Constants}
With the recent publication of perturbation constants for the \upperstate\ \textit{v}=4 and \textit{v}=6 levels by Bornhauser \etal\ (Table \ref{TABperts}), there was an opportunity to improve the molecular constants reported by Tanabashi \etal\ (Tables 3 and 4 in ref~\citep{2007Tanabashi-a}). In their calculation of molecular constants in 2007, Tanabashi \etal\ included lines from several other studies. For nine bands up to (3,4), gaps in observations were filled in by using lines from Prasad and Bernath~\citep{1994Prasad-a}. High resolution measurements of the (0,1) band by Curtis and Sarre~\citep{1985Curtis-a} were included, as were cross transitions ($\Delta\Omega\neq$0) from Suzuki \etal\ \citep{1985Suzuki-a} for the (1,0) band. Some cross transitions were also observed by Curtis and Sarre, and these are particularly useful for the accurate calculation of the spin-orbit coupling and $\Lambda$-doubling constants. All lines from Tanabashi and Amano~\citep{2002Tanabashi-a} for the (5,7), (6,8) and (7,9) bands were also included.

Prasad and Bernath also calculated molecular constants, and included in their fit all lines from Curtis and Sarre, Suzuki \etal\ and Amiot (for the (0,0) band). Our recalculation is based mainly on that of Tanabashi \etal, and also this fit by Prasad and Bernath. An explanation of the specific differences is presented below.

The computer program \emph{PGOPHER}~\citep{2010Western-Misc-a}, written by C. M. Western (University of Bristol) was used to recalculate the molecular constants, with the inclusion of the \textit{v\primed}=4 and \textit{v\primed}=6 perturbations, using the standard $N^2$ Hamiltonian for a $^3\Pi$ state~\citep{1979Brown-a, 1994Hirota-a}. A global least squares fit was performed including all lines from Tanabashi \etal, Tanabashi and Amano, Prasad and Bernath, Curtis and Sarre, Suzuki \etal, Lloyd and Ewart~\citep{1999Lloyd-a} (for the (0,0) band) and the two deperturbation studies by Bornhauser \etal~\citep{2010Bornhauser-a, 2011Bornhauser-a}.

The weights for the lines from Tanabashi \etal\ (including those from Tanabashi and Amano) were unchanged here, except for those involving the \textit{v\primed}=4 and \textit{v\primed}=6 states. These had mostly been deweighted, and were weighted more strongly in this study as the perturbations had been included in the fit. In their calculation of the perturbation constants, Bornhauser \etal\ observed lines involving \textit{J}\primed=1-6, 10-12 and 17-23 for the \textit{v\primed}=6 state, and \textit{J}\primed=4-14 for the \textit{v\primed}=6 state. Lines involving these \textit{J} levels were weighted highly, and other line weights were decreased with greater difference between \textit{J}\primed\ and these ranges. The actual lines observed by Bornhauser \etal\ were weighted similarly to those of Tanabashi \etal\ for the same bands.

The five remaining sets of lines were treated as follows. The sets from Prasad and Bernath (two sets), Curtis and Sarre and Suzuki \etal\ were given the same weights as in the fit performed by Prasad and Bernath. Lloyd and Ewart lines were assigned weights to be similar to those of Tanabashi \etal\ for the (0,0) band. To ensure that all lines were on the same wavenumber scale, transitions from these five sets were then compared to matching Tanabashi \etal\ transitions, and a weighted average wavenumber difference (one for each set) using matching lines was calculated, based on the assigned weights. This was added to all of the lines from each set as a wavenumber offset, to compensate for any systematic differences between studies. In their fit, Tanabashi \etal\ deweighted many lines due to the extensive perturbations, and those lines were also deweighted here if they were present in these five sets. This process excluded approximately 11\% of these lines. To further decrease the possibility of using any misassigned lines in the fit, any line whose wavenumber differed from a matching Tanabashi \etal\ line by more than 0.03 \wavenumbers\ was deweighted, excluding a further $\sim$6\%. A preliminary fit was then performed to obtain calculated values of each transition. Lines that had not been matched to Tanabashi \etal\ transitions were then deweighted if their observed-calculated values, as a result of this fit, were greater than 0.03 \wavenumbers.

A final global weighted least squares fit was performed, in which all reported molecular constants for the \lowerstate\ and \upperstate\ were floated, except for $A_D$ for \textit{v}\primed=8, 9 and 10. These were fixed at a value based on those calculated for the lower levels to obtain a good fit. The updated molecular constants are shown in Tables \ref{TABnewConstantsd} and \ref{TABnewConstantsa}. The magnitudes of the perturbation constants reported by Bornhauser \etal\ were also floated to improve the fit, and both the previous and changed values are shown in Table \ref{TABperts}.

\section{Calculation of Line Intensities}
The intensities of the rovibronic transitions are reported here as both Einstein \textit{A} values and oscillator strengths (\emph{f}-values).

Einstein \textit{A} values are calculated with the equation \citep{2005Bernath-Book-a}

\vspace{-0.5cm}
\begin{eqnarray}\label{EQNEinA}
A_{J^\prime\rightarrow J^{\prime\prime}} & = & \frac{16\pi \nu^3S_{J^{\prime\prime}}^{\Delta J}}{3\epsilon_0 hc^3(2J^\prime+1)}\langle \psi_{v^\prime J^\prime}|R_e(r)|\psi_{v^{\prime\prime}J^{\prime\prime}}\rangle^2\\
 & = & 3.136\ 189\ 32\ \times 10^{-7} \frac{\bar{\nu}^3S_{J^{\prime\prime}}^{\Delta J}}{(2J^\prime+1)}|\langle \psi_{v^\prime J^\prime}|R_e(r)|\psi_{v^{\prime\prime}J^{\prime\prime}}\rangle|^2,
\end{eqnarray}

where $S_{J^{\prime\prime}}^{\Delta J}$ is the H\"{o}nl-London factor and $|\langle \psi_{v^\prime J^\prime}|R_e(r)|\psi_{v^{\prime\prime}J^{\prime\prime}}\rangle|$ is the transition dipole moment (TDM) matrix element, $A_{J^\prime\rightarrow J^{\prime\prime}}$ is in s\super{-1}, $\bar{\nu}$ in cm\super{-1} and R\sub{e} in debye. These have been converted into \emph{f}-values using the equation
\vspace{-0.5cm}
\begin{eqnarray}\label{EQNAtof}
f_{J^\prime\leftarrow J^{\prime\prime}} & = & \frac{m_e \epsilon_0 c (2J^{\prime}+1)}{2\pi e^2 (100\bar{\nu})^2 (2J^{\prime\prime}+1) } A_{J^\prime\rightarrow J^{\prime\prime}}\\
 & = & {1.499\ 193\ 68\ }\frac{1}{\bar{\nu}^2} \frac{(2J^{\prime}+1)}{(2J^{\prime\prime}+1) }A_{J^\prime\rightarrow J^{\prime\prime}}.
\end{eqnarray}

Band intensities are reported as \Avv\ values, and can be converted from these to \fvv\ values using the equation:
\begin{equation}\label{EQNAvvtofvv}
f_{v^\prime\leftarrow v^{\prime\prime}} = {1.499\ 193\ 68\ }\frac{1}{\bar{\nu}^2} A_{v^\prime\rightarrow v^{\prime\prime}}
\end{equation}
where $\bar{\nu}$ is the average wavenumber for the band \citep{1983Larsson-a}.

\textit{PGOPHER} was used to calculate Einstein \textit{A} values. It is able to calculate the necessary rotational TDMs and H\"{o}nl-London factors (Eq. \ref{EQNEinA}) for several types of electronic transitions, including \tPI-\tPI, if provided with a set of molecular constants and band strengths for each vibrational band. 

 For a diatomic molecule, the wavefunctions $\psi_{vJ}$ used in the calculation of the TDM can be described as a one-dimensional function of internuclear distance~\citep{2005Bernath-Book-a}. Rotationless TDMs were calculated using the computer program \emph{LEVEL}~\citep{2007LeRoy-Report-a}, written by R. J. Le Roy, which is able to calculate eigenfunctions and eigenvalues by solving the one-dimensional Schr\"{o}dinger equation for diatomic molecules. \emph{LEVEL} is able to calculate TDMs for rotational levels above \textit{J}=0, but it assumes a singlet-singlet transition. For this reason, a single TDM (for Q(0)) for each vibrational band was taken from \emph{LEVEL} and input into \emph{PGOPHER}.

\emph{LEVEL} must be provided with a potential energy curve, \emph{V}(\emph{r}), and an electronic TDM, both as a function of internuclear distance.

\subsection{Electronic Transition Dipole Moment}
Our calculation of the electronic TDM of the Swan system has been reported previously~\citep{2007Kokkin-a,2007Schmidt-a,2009Nakajima-a}, with the results shown in Table \ref{TABeTDM} and Figure \ref{FIGeTDM}. A brief description is given here. Wavefunctions were computed using the multi-reference configuration interaction (MRCI) method~\citep{1988Werner-a,1988Knowles-a}, whereby all single and double excitations from a complete active space self-consistent field (CASSCF)\citep{1985Werner-a,1985Knowles-a} reference state are included in the MRCI wave functions. The active space included all molecular orbitals (MO) arising from the C atoms' $2s$ and $2p$ valence orbitals. The basis set is the augmented correlation-consistent polarized aug-cc-pV6Z set of Dunning and co-workers~\citep{1989Dunning-a,1992Kendall-a,1995Woon-a,1996Wilson-a} and de Jong \emph{et al} .\citep{2001deJong-a} Core and core-valence (CV) correlation corrections were obtained using the aug-cc-pCVQZ basis set~\citep{1989Dunning-a,1992Kendall-a,1995Woon-a}. Scalar relativistic energy corrections (Rel) were evaluated via the Douglas-Kroll-Hess approach~\citep{1974Douglas-a,1985Hess-a,1986Hess-a}, in conjunction with the appropriate cc-pVQZ basis sets. The quantum chemical calculations were carried out using the MOLPRO2006.1 program~\citep{2006Werner-Misc-a}.

\subsection{Potential Energy Curves}

\begin{figure}
    \centering
  \includegraphics{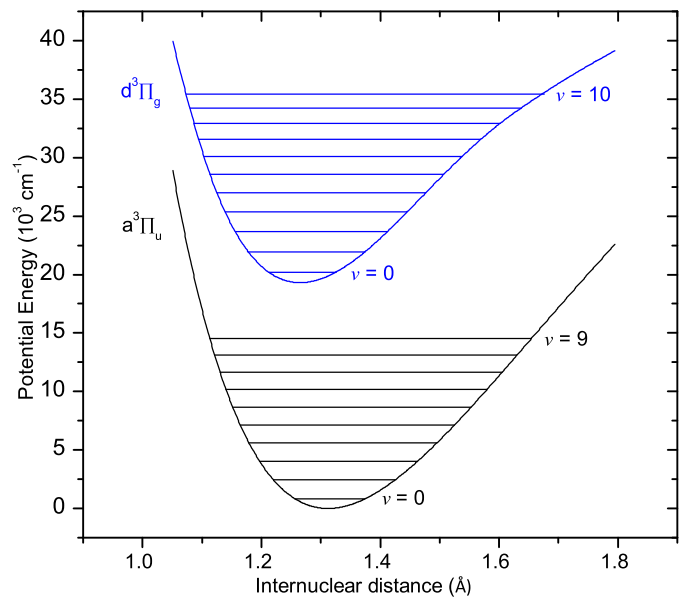}
  \caption{The potential energy curves of the \CC\ Swan System.}\label{FIGpotentials}
\end{figure}

\begin{figure}
\begin{center}
  \includegraphics[width=9cm]{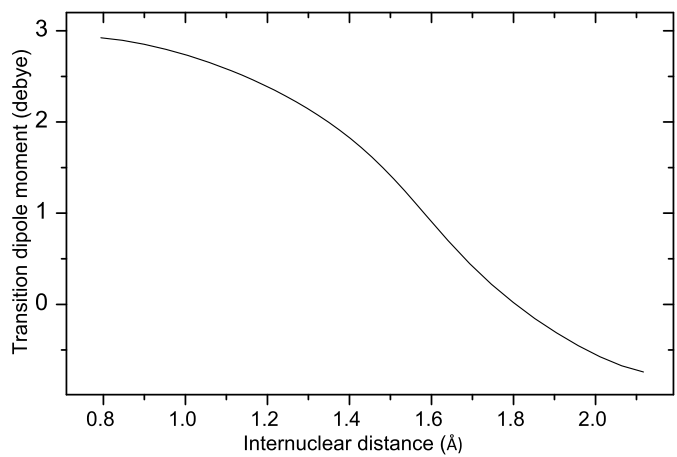}
  \caption[width=9cm]{The electronic transition dipole moment of the \CC\ Swan system}\label{FIGeTDM}
\end{center}
\end{figure}

 The potentials \emph{V}(\emph{r}) (Fig. \ref{FIGpotentials}) were calculated using the computer program \emph{RKR1}~\citep{2004LeRoy-Report-a}, which utilizes the first-order semiclassical Rydberg-Klein-Rees procedure~\citep{1932Rydberg-a,1933Rydberg-a,1932Klein-a,1947Rees-a} to determine a set of classical turning points for each potential, using equilibrium expansion constants $\omega_{e}$, $\omega_{e}x_{e}$, $\omega_{e}y_{e}$, $\omega_{e}z_{e}$, $\alpha_{e}$, $\gamma_{e}$ and $\delta_{e}$ for \Bv\ and \Gv. Values for these constants were calculated (Table \ref{TABconst}) in a weighted least squares fit using the energy level expressions for a vibrating rotator,
\begin{equation}\label{Gv}
G(v) = \omega _{e}(v+\frac{1}{2}) - \omega _{e}x_{e}(v+\frac{1}{2})^2 + \omega _{e}y_{e}(v+\frac{1}{2})^3 + \omega _{e}z_{e}(v+\frac{1}{2})^4
\end{equation}
and
\begin{equation}\label{Bv}
B_{v} = B_e -\alpha_{e}(v+\frac{1}{2}) + \gamma _{e}(v+\frac{1}{2})^2 + \delta _{e}(v+\frac{1}{2})^3,
\end{equation}

and the updated \Bv\ and \Gv\ values in Tables \ref{TABnewConstantsd} and \ref{TABnewConstantsa}. The weightings of the vibrational levels were based on the standard deviation of \Bv\ and \Gv\ values from the $PGOPHER$ line position fit.

\subsection{Vibrational Band Intensities}

\Avv\ values for each vibrational band were also calculated, and are shown in Table \ref{TABAvv}. They were calculated as the sum of all single rotational Einstein \textit{A} values for possible transitions within the relevant band from the \textit{J}\primed=1, $\Omega$\primed=0 level. These were converted into \fvv\ values using Equation \ref{EQNAvvtofvv}.

\section{Analysis and Discussion}
Our final line list including positions, \emph{f}-values and Einstein \textit{A} values, is available at http://bernath.uwaterloo.ca/download/AutoIndex.php?dir=/C2/ (C2SwanLineList2012.txt). Calculated line positions and lower state energies are included for all lines, and observed positions are present when available. Line intensities are reported as both Einstein \textit{A} values and \textit{f}-values. Positions and intensities were calculated for all possible bands involving the observed vibrational levels (i.e. up to (0,9) and (10,0)).

\emph{PGOPHER} was also used for the purpose of validation, as it is able to calculate and plot spectra based on its line list, which can be compared to the observed spectrum. In all of the spectra shown, a constant Gaussian instrument function was added to best match the observed broadening. The experimental procedure that Tanabashi \emph{et al}. used involved observing \CC\ emission from a microwave discharge in a flow of acetylene (C\sub{2}H\sub{2}) diluted in argon through a discharge tube. In such a system the molecular vibration, and to a lesser extent the rotation, will not be at thermal equilibrium. For this reason, the rotational and vibrational temperatures of the simulation were also adjusted for best agreement. Two spectra were recorded, one for the $\Delta$\emph{v}=-1 to +2 sequences and another for the $\Delta$\emph{v}=-2 and -3 sequences. Rotational and vibrational temperatures of 1140 K and 6800 K, and 940 K and 5000 K were used for the $\Delta$\emph{v}=-1 to +2 and $\Delta$\emph{v}=-2 to -3 spectra, respectively. The final parameter that had to be added manually was a linear scaling factor, as the y-axis units of the recorded spectrum are arbitrary. This value could not be kept constant during the production of each figure given below (Figs. \ref{fig(0,0)} to \ref{fig(2,0)}). This is due to the presence of an instrument response function, which cannot be corrected for at this point.

There are numerous perturbations in the \upperstate\ state, which have caused many of the line positions calculated by \emph{PGOPHER} to be slightly inaccurate, and in turn have also had a small effect on the reported intensities. With the inclusion of the perturbation constants for the \textit{v\primed}=4 and \textit{v\primed}=6 levels, the average error for lines involving those upper levels is improved from 0.203 \wavenumbers\ to 0.057 \wavenumbers\, and 0.569 \wavenumbers\ to 0.038 \wavenumbers, respectively. The first values were calculated using the molecular constants of Tanabashi \etal\, and the second using those in Tables \ref{TABnewConstantsd} and \ref{TABnewConstantsa}, excluding any lines that had been heavily deweighted in the final fit. A more detailed description of the observed perturbations is available in ref~\citep{2007Tanabashi-a}.

The spectra match very well for the lower vibrational and rotational levels; most of the inaccuracies mentioned are present in the higher vibrational and rotational levels. Three small sections of the spectra are shown in Figures \ref{fig(0,0)} to \ref{fig(2,0)} that are typical of the rest of the range.

For further validation, lifetimes of vibrational levels have been calculated and are compared to previous theoretical and experimental results in Table \ref{TABLifetimes}. For each upper vibrational level, lifetimes were calculated as the reciprocal of the sum of the Einstein \textit{A} values for all possible transitions from the \textit{J}\primed=1, $\Omega$\primed=0 level. Good agreement is shown with both sets of data. The theoretical values of Schmidt and Bacskay include transitions to the c$^3\Sigma\rm{_g^+}$ state. They state that this system contributes 3-4\% to their radiative lifetimes, and if this is taken into account, excellent agreement with our values is shown. Our \Avv\ values were converted into \fvv\ values for comparison with those of Schmidt \etal\ (for up to \textit{v}\primed=5 and \textit{v}\primed\primed=5) \citep{2007Schmidt-a}, using Equation \ref{EQNAvvtofvv} and the wavenumber of the P(0) transition as the band wavenumber. It should be noted that slightly different \fvv\ values would be obtained with a different choice of band wavenumber. Excellent agreement is shown for most bands, however some of the higher vibrational bands disagree by up to $\sim$60\%, as shown in Table \ref{TABfvv}.

\section{Conclusion}
Many perturbations are present in the \upperstate\ state, and only those shown in Table \ref{TABperts} (involving the \upperstate, \textit{v}=4 and \textit{v}=6 levels) have been accounted for in these calculations. While many of the lines in the new line list do not match experiment precisely, the positions and intensities reported here are an improvement on previously available data, where results have been based on the partly incorrect assignments made by Phillips and Davis~\citep{1968Phillips-Book-a}. The calculated vibrational level lifetimes show good agreement with experimental and theoretical studies. The line list produced is an improvement over what is currently available, and will be of use to astronomers, materials and combustion scientists in the analysis of the \CC\ Swan system.

\section{Acknowledgements}

Support for this work was provided by a Research Project Grant from the Leverhulme Trust and a Department of Chemistry (University of York) studentship.



\begin{table}

\vspace*{3mm}
\centering
  \begin{threeparttable}
  \caption{Perturbation constants\tnote{a} for the \upperstate\, \textit{v}\primed=4 and \textit{v}\primed=6 levels of the \CC\ Swan system. Those of Bornhauser \etal\ \citep{2010Bornhauser-a, 2011Bornhauser-a}, and those resulting from the fit of all molecular constants are reported.}
    \begin{tabular}{lr@{}lr@{}l}
        \hline\noalign{\smallskip}
        Parameter &\multicolumn{2}{c}{Bornhauser \etal\ value} &\multicolumn{2}{c}{Value from fit}\\
        \hline\noalign{\smallskip}
        $\langle$\upperstate, \textit{v}\primed=4$|$H\sub{SO}$|$\bSigma, \textit{v}\primed=16$\rangle$    & -0.&6401(86)\tnote{b}  & -0.&6147(59)  \\
        $\langle$\upperstate, \textit{v}\primed=4$|$BL\sub{+}$|$\bSigma, \textit{v}\primed=16$\rangle$    &  0.&24737(61) &  0.&24869(21) \\
        $\langle$\upperstate, \textit{v}\primed=6$|$H\sub{SO}$|$\bSigma, \textit{v}\primed=19$\rangle$    & 0.&7855(110)  & 0.&7417(82)   \\
        $\langle$\upperstate, \textit{v}\primed=6$|$BL\sub{+}$|$\bSigma, \textit{v}\primed=19$\rangle$    & 0.&31192(37)  & 0.&31123(12)  \\
        $\langle$\upperstate, \textit{v}\primed=6$|$H\sub{SO}$|$\OnePi$\rangle$ \ \ \ \ \ \ \ \ \ \       & 4.&6220(88)   & 4.&6150(94)   \\ [1ex]
        \hline
    \label{TABperts}
    \end{tabular}
    \vspace*{-5mm}
    \begin{tablenotes}
       \begin{footnotesize}
       \item[a] Numbers in parentheses indicate one standard deviation to the last significant digits of the constants.
       \item[b] {Please note that these values were floated to improve the fit, but the studies of Bornhauser \etal\ were directly aimed at calcualting these constants.}
       \end{footnotesize}
    \end{tablenotes}
  \end{threeparttable}

\end{table}

\begin{table}

\begin{center}
\scalebox{0.75}{
\hspace*{-4.19cm}
\begin{threeparttable}
\renewcommand{\captionfont}{\large}
\caption{Updated molecular constants\tnote{a} for the \upperstate\ state of  the \CC\ Swan system (in cm$^{-1}$).}

\begin{tabular}{cr@{}lr@{}lr@{}lr@{}lr@{}lr@{}lr@{}lr@{}lr@{}l}
\hline\noalign{\smallskip}
{$v$} &\multicolumn{2}{c}{$T_v$} &\multicolumn{2}{c}{$A$} &\multicolumn{2}{c}{$A_D$} &\multicolumn{2}{c}{$B$} &\multicolumn{2}{c}{$D$$\times$10$^6$} &\multicolumn{2}{c}{$\lambda$} &\multicolumn{2}{c}{$o$} &\multicolumn{2}{c}{$p$} &\multicolumn{2}{c}{$q$} \\
\hline\noalign{\smallskip}
0     &   19378.&46833(33) &   -14.&00085(41)  &   0.&0004828(56)   &   1.&7455648(28)  &   6.&8203(10)     &  0.&03313(30)   &   0.&61041(34)    &   0.&003963(29)   &   -0.&0007780(28) \\
1\tnote{b}     &   21132.&14979(16) &   -13.&87448(32)  &   0.&0005533(55)   &   1.&7254068(35)  &   7.&0202(50)     &  0.&02978(25)   &   0.&61704(23)    &   0.&004117(29)   &   -0.&0008200(29) \\
2     &   22848.&3971(14)  &   -13.&8216(15)   &   0.&000478(29)    &   1.&704476(14)   &   7.&281(14)      &  0.&0199(27)    &   0.&6141(21)     &   0.&00643(26)    &   -0.&0008593(95) \\
3     &   24524.&2219(13)  &   -13.&5311(20)   &   0.&000595(15)    &   1.&681467(11)   &   7.&502(16)      &  0.&0457(19)    &   0.&5727(19)     &   0.&00517(12)    &   -0.&0008549(58) \\
4     &   26155.&0539(28)  &   -13.&3521(58)   &   0.&000657(59)    &   1.&656891(27)   &   7.&681(36)      &  0.&0394(62)    &   0.&5833(48)     &   0.&00679(41)    &   -0.&000906(29)  \\
5     &   27735.&6812(38)  &   -13.&0309(67)   &   0.&000528(35)    &   1.&630278(23)   &   8.&733(28)      &  0.&0755(45)    &   0.&5634(47)     &   0.&00519(27)    &   -0.&000877(11)  \\
6     &   29259.&3610(29)  &   -12.&7946(83)   &   0.&000830(47)    &   1.&599907(23)   &   9.&064(31)      &  0.&0574(59)    &   0.&5485(61)     &   0.&00724(38)    &   -0.&000975(16)  \\
7     &   30717.&9213(51)  &   -12.&3399(92)   &   0.&000878(45)    &   1.&565926(35)   &   9.&780(68)      &  0.&0890(33)    &   0.&5123(67)     &   0.&00874(48)    &    0.&001293(24)  \\
8     &   32102.&668(14)   &   -12.&100(15)    &   0.&000556(fixed) &   1.&52676(20)    &   9.&62(63)       &  0.&038(19)     &   0.&490(15)      &   0.&0047(14)     &   -0.&00089(14)   \\
9     &   33406.&256(15)   &   -11.&676(27)    &   0.&000556(fixed) &   1.&485411(82)    &   11.&09(11)      &  0.&173(21)     &   0.&451(21)      &   0.&0087(12)     &   -0.&002158(41)  \\
10    &   34626.&8079(74)  &   -11.&263(13)    &   0.&000556(fixed) &   1.&441006(56)   &   12.&547(78)     &  0.&113(11)     &   0.&357(10)      &   0.&00676(75)    &   -0.&001127(33)  \\
\hline\noalign{\smallskip}
\label{TABnewConstantsd}
\end{tabular}
    \vspace*{-5mm}
    \begin{tablenotes}
       \begin{large}
       \item[a] Numbers in parentheses indicate one standard deviation to the last significant digits of the constants.
       \item[b] In addition, $H$ = 2.16(19)$\times$10$^{-11}$ for \textit{v}=1 was used to obtain a good fit.
       \end{large}
    \end{tablenotes}
\end{threeparttable}}
\end{center}
\end{table}

\begin{table}

\begin{center}
\scalebox{0.75}{
\hspace*{-4.2cm}
\begin{threeparttable}
\renewcommand{\captionfont}{\large}
\caption{Updated molecular constants\tnote{a} for the \lowerstate\ state of  the \CC\ Swan system (in cm$^{-1}$).}

\begin{tabular}{cr@{}lr@{}lr@{}lr@{}lr@{}lr@{}lr@{}lr@{}lr@{}l}
\hline\noalign{\smallskip}
{$v$} &\multicolumn{2}{c}{$T_v$} &\multicolumn{2}{c}{$A$} &\multicolumn{2}{c}{$A_D$} &\multicolumn{2}{c}{$B$} &\multicolumn{2}{c}{$D$$\times$10$^6$} &\multicolumn{2}{c}{$\lambda$} &\multicolumn{2}{c}{$o$} &\multicolumn{2}{c}{$p$} &\multicolumn{2}{c}{$q$} \\
\hline\noalign{\smallskip}
0\tnote{b}     &       0&          &   -15.&26930(28)  &   0.&0002441(47)  &   1.&6240458(29)  &   6.&4515(12)      &  -0.&15446(23)   &   0.&67526(23)    &   0.&002499(30)   &   -0.&0005303(29) \\
1     &   1618.&02331(35) &   -15.&25156(40)  &   0.&0002063(49)  &   1.&6074247(29)  &   6.&4433(14)      &  -0.&15364(33)   &   0.&66991(33)    &   0.&002693(30)   &   -0.&0005799(28) \\
2     &   3212.&72927(63) &   -15.&2326(10)   &   0.&0001679(64)  &   1.&5907495(41)  &   6.&4530(29)      &  -0.&15228(80)   &   0.&66251(91)    &   0.&003089(51)   &   -0.&0006484(32) \\
3     &   4784.&1038(21)  &   -15.&2148(27)   &   0.&000040(28)   &   1.&573971(16)   &   6.&377(16)       &  -0.&1510(41)    &   0.&6565(36)     &   0.&00516(29)    &   -0.&000653(11)  \\
4     &   6332.&1379(34)  &   -15.&2020(44)   &   0.&000161(26)   &   1.&557173(21)   &   6.&449(27)       &  -0.&1520(50)    &   0.&6504(46)     &   0.&00520(26)    &   -0.&000892(11)  \\
5     &   7856.&8251(22)  &   -15.&1941(34)   &   -0.&0000051(19) &   1.&540145(16)   &   6.&353(24)       &  -0.&1520(31)    &   0.&6389(28)     &   0.&00637(18)    &   -0.&0012365(83) \\
6     &   9358.&1689(30)  &   -15.&1631(48)   &   0.&000178(26)   &   1.&523474(21)   &   6.&151(28)       &  -0.&1496(41)    &   0.&6492(39)     &   0.&00375(25)    &   -0.&000652(11)  \\
7     &   10836.&1483(65) &   -15.&0917(94)   &   0.&00058(20)    &   1.&50895(17)    &   3.&41(61)        &  -0.&1537(69)    &   0.&654(10)      &  -0.&0288(19)     &    0.&00686(36)   \\
8     &   12290.&8297(65) &   -15.&142(11)    &   -0.&000068(59)  &   1.&488368(50)   &   4.&648(86)       &  -0.&1435(90)    &   0.&6382(81)     &   0.&01246(56)    &   -0.&002132(29)  \\
9     &   13722.&1088(51) &   -15.&0901(84)   &   0.&000468(39)   &   1.&472699(34)   &   5.&795(64)       &  -0.&1677(33)    &   0.&6508(67)     &   0.&00268(50)    &   -0.&000214(28)  \\
\hline\noalign{\smallskip}
\label{TABnewConstantsa}
\end{tabular}
    \vspace*{-5mm}
    \begin{tablenotes}
       \begin{large}
       \item[a] Numbers in parentheses indicate one standard deviation to the last significant digits of the constants.
       \item[b] In addition, $H$ = 6.87(11)$\times$10$^{-12}$, $o_D$ = -7.14(74)$\times$10$^{-6}$, $p_D$ = 3.0(19)$\times$10$^{-8}$ and $q_D$ = 1.000(36)$\times$10$^{-8}$ were used for \textit{v}=0 to obtain a good fit.
       \end{large}
    \end{tablenotes}
\end{threeparttable}}
\end{center}
\end{table}

\begin{table}
\renewcommand{\captionfont}{\normalsize}
\caption{Equilibrium molecular constants$\rm{^a}$ for the \CC\ Swan system.}
\label{TABconst}
\centering
\begin{threeparttable}
    \begin{tabular}{cr@{}lr@{}l}
        \hline\noalign{\smallskip}
        Constant &\multicolumn{2}{c}{d$^3\Pi\rm{_g}$}  &\multicolumn{2}{c}{a$^3\Pi\rm{_u}$} \\
        \hline\noalign{\smallskip}
        $\omega${\sub{e}}                   & 1788.&52(33)        & 1641.&3451(38)    \\
        $\omega${\sub{e}}\emph{x}\sub{e}    & 16.&92(33)          & 11.&6583(14)      \\
        $\omega${\sub{e}}\emph{y}\sub{e}    & -0.&250(38)         & -0.&000888(122)     \\
        $\omega${\sub{e}}\emph{z}\sub{e}    & -0.&0401(22)        &&-    \\
        B\sub{e}                            &  1.&755410(81)       & 1.&632343(46)    \\
        $\alpha${\sub{e}}                   & -0.&01960(12)       & -1.&6569(40)   \\
        $\gamma${\sub{e}}                   & -0.&000141(38)      & -0.&0000294(61) \\
        $\delta${\sub{e}}                   & -0.&0000809(31)     &&- \\ [1ex]
        \hline\noalign{\smallskip}
    \end{tabular}
    \vspace*{-2mm}
    \begin{tablenotes}
       \begin{footnotesize}
        \item[a] Numbers in parentheses indicate one standard deviation to the last significant digits of the constants.
       \end{footnotesize}
    \end{tablenotes}

\end{threeparttable}
\end{table}

\begin{table}
\hspace{-1cm}
\begin{threeparttable}

      \caption{Calculated transition dipole moment function for the \CC\ Swan system.}
    \begin{tabular}{c c | c c | c c | c r@{}l}

        \hline\noalign{\smallskip}
        $r$ & $R_e$ & $r$ & $R_e$ & $r$ & $R_e$ & $r$ &\multicolumn{2}{c}{$R_e$}\\
        (au) & (au) & (au) & (au) & (au) & (au) & (au) &\multicolumn{2}{c}{(au)}\\
        \hline\noalign{\smallskip}
        1.50 &	1.150484941	&	2.25	&	0.947233687 &   2.70	&	0.676887334	&   3.30 &	0.&086826219  \\
        1.60 &	1.139421001	&	2.30    &	0.924302460	&	2.75	&	0.634819590	&	3.40 &	0.&008468806  \\
        1.70 &	1.122420868	&	2.35	&	0.899942917	&	2.80	&	0.589585866	&	3.50 &	-0.&060876772 \\
        1.80 &	1.100556366	&	2.40	&	0.874060543	&	2.85	&	0.541254759	&	3.60 &	-0.&122739863 \\
        1.90 &	1.074421199	&	2.45	&	0.846525942	&	2.90	&	0.490200083	&	3.70 &	-0.&177872176 \\
        2.00 &	1.043723590	&	2.50	&	0.817168296	&	2.95	&	0.437092215	&	3.80 &	-0.&225903802 \\
        2.10 &	1.008582913	&	2.55	&	0.785779108	&	3.00	&	0.382865673	&	3.90 &	-0.&264852598 \\
        2.15 &	0.989349200	&	2.60	&	0.752116932	&	3.10	&	0.275576227 &	4.00 &	-0.&291460152 \\
        2.20 &	0.968873791	&	2.65	&	0.715908977 &   3.20	&	0.175777762 \\ [1ex]
        \hline\noalign{\smallskip}
        \label{TABeTDM}
    \end{tabular}
\end{threeparttable}

\end{table}

\begin{table}
\caption{Lifetimes of vibrational levels of the \CC\ \upperstate\ state.}\label{TABLifetimes}
\centering
\begin{threeparttable}

    \begin{tabular}{ l r@{}l c r@{}l c }
        \hline\noalign{\smallskip}
        \textit{v}\primed &\multicolumn{2}{l}{Our value} & Theoretical\tnote{a}\ \citep{2007Schmidt-a}  &\multicolumn{2}{c}{Expt.~\citep{1988Naulin-a}}  & Expt.~\citep{1986Bauer-a}\\
         &\multicolumn{2}{l}{(ns)} & (ns) &\multicolumn{2}{c}{(ns)} & (ns)\\
        \hline\noalign{\smallskip}
        0    & 98.&0  & 95.1   & 101.8$\pm$&4.2 & 106$\pm$15 \\
        1    & 99.&8  & 96.7   & 96.7$\pm$&5.2  & 105$\pm$15 \\
        2    & 102.&4 & 99.1  & 104.0$\pm$&17 & \\
        3    & 105.&8 & 102  \\
        4    & 110.&4 & 107  \\
        5    & 116.&9 & 113  \\ [1ex]
        \hline\noalign{\smallskip}
    \end{tabular}
    \vspace*{-2mm}
    \begin{tablenotes}
        \begin{footnotesize}
            \item[a] The theoretical values (of Schmidt and Bacskay) include transitions to the c$^3\Sigma\rm{_g^+}$ state. They state that this system contributes 3-4\% to their radiative lifetimes. If this is taken into account, excellent agreement with our values is shown.
        \end{footnotesize}
    \end{tablenotes}

\end{threeparttable}
\end{table}

\begin{table}
\begin{center}
\scalebox{0.75}
{
\hspace*{-4.2cm}
\begin{threeparttable}
\renewcommand{\captionfont}{\large}
\caption{Einstein \textit{A}$_{v^{\prime}}$$_{v^{\prime\prime}}$ values\tnote{a} of the \CC\ Swan system.}
\begin{tabular}{l r@{}l r@{}l r@{}l r@{}l r@{}l r@{}l r@{}l r@{}l r@{}l r@{}l r@{}l }
        \hline\noalign{\smallskip}
        &\multicolumn{22}{c}{\textit{v}\primed} \\
        \cline{2-23}\noalign{\smallskip}
        \textit{v}\primed\primed &\multicolumn{2}{c}{0} &\multicolumn{2}{c}{1}  &\multicolumn{2}{c}{2}  &\multicolumn{2}{c}{3}  &\multicolumn{2}{c}{4}  &\multicolumn{2}{c}{5}  &\multicolumn{2}{c}{6}  &\multicolumn{2}{c}{7}  &\multicolumn{2}{c}{8}  &\multicolumn{2}{c}{9}  &\multicolumn{2}{c}{10} \\
        \hline\noalign{\smallskip}
0 &  7.&626 (+6)  &  2.&815 (+6)  &  2.&811 (+5)  &  4.&386 (+3)  &  1.&830 (+2)  &  2.&641 (+1)  &  4.&709 (-1)  &  1.&616 (-1)  &  4.&647 (-2)  &  1.&136 (-4)  &  1.&298 (-2)  \\
1 &  2.&135 (+6)  &  3.&427 (+6)  &  4.&074 (+6)  &  6.&452 (+5)  &  9.&222 (+3)  &  1.&341 (+3)  &  9.&819 (+1)  &  7.&970 (+0)  &  3.&257 (-1)  &  2.&280 (-1)  &  4.&324 (-2)  \\
2 &  3.&833 (+5)  &  2.&746 (+6)  &  1.&272 (+6)  &  4.&433 (+6)  &  9.&730 (+5)  &  9.&006 (+3)  &  4.&910 (+3)  &  1.&818 (+2)  &  5.&525 (+1)  &  1.&885 (-1)  &  8.&546 (-1)  \\
3 &  5.&588 (+4)  &  8.&279 (+5)  &  2.&564 (+6)  &  3.&279 (+5)  &  4.&337 (+6)  &  1.&201 (+6)  &  2.&624 (+3)  &  1.&430 (+4)  &  7.&755 (+1)  &  2.&377 (+2)  &  2.&902 (+0)  \\
4 &  7.&225 (+3)  &  1.&709 (+5)  &  1.&171 (+6)  &  2.&057 (+6)  &  2.&796 (+4)  &  4.&082 (+6)  &  1.&211 (+6)  &  1.&463 (+3)  &  2.&886 (+4)  &  1.&613 (+2)  &  5.&789 (+2)  \\
5 &  8.&589 (+2)  &  2.&885 (+4)  &  3.&212 (+5)  &  1.&355 (+6)  &  1.&496 (+6)  &  1.&131 (+4)  &  3.&582 (+6)  &  1.&228 (+6)  &  3.&360 (+4)  &  4.&195 (+4)  &  3.&517 (+3)  \\
6 &  9.&554 (+1)  &  4.&284 (+3)  &  6.&785 (+4)  &  4.&745 (+5)  &  1.&385 (+6)  &  1.&019 (+6)  &  6.&585 (+4)  &  3.&648 (+6)  &  9.&982 (+5)  &  1.&426 (+5)  &  3.&709 (+4)  \\
7 &  1.&006 (+1)  &  5.&771 (+2)  &  1.&220 (+4)  &  1.&216 (+5)  &  6.&015 (+5)  &  1.&306 (+6)  &  6.&220 (+5)  &  1.&047 (+5)  &  3.&561 (+6)  &  6.&228 (+5)  &  3.&548 (+5)  \\
8 &  1.&001 (+0)  &  7.&183 (+1)  &  1.&950 (+3)  &  2.&582 (+4)  &  1.&828 (+5)  &  6.&851 (+5)  &  1.&092 (+6)  &  4.&253 (+5)  &  8.&571 (+4)  &  3.&500 (+6)  &  2.&098 (+5)  \\
9 &  9.&107 (-2)  &  8.&362 (+0)  &  2.&826 (+2)  &  4.&801 (+3)  &  4.&491 (+4)  &  2.&419 (+5)  &  6.&728 (+5)  &  1.&008 (+6)  &  2.&739 (+5)  &  3.&371 (+4)  &  3.&325 (+6)  \\ [1ex]
        \hline\noalign{\smallskip}
        \label{TABAvv}
    \end{tabular}
    \vspace*{-5mm}
    \begin{tablenotes}
        \begin{large}
            \item[a] The numbers in parentheses indicate the exponent.
        \end{large}
    \end{tablenotes}
\end{threeparttable}
}
\end{center}

\end{table}

\begin{table}

\begin{center}
\scalebox{0.65}{
\hspace*{-4.2cm}
\begin{threeparttable}
\renewcommand{\captionfont}{\large}
\caption{$f_{v^{\prime}}$$_{v^{\prime\prime}}$ values\tnote{a} of the \CC\ Swan system (a), compared to those of Schmidt \etal\ \citep{2007Schmidt-a} (b).}
   \begin{tabular}{l c r@{}l r@{}l c r@{}l r@{}l c r@{}l r@{}l c r@{}l r@{}l c r@{}l r@{}l c r@{}l r@{}l}
        \hline\noalign{\smallskip}
        & &\multicolumn{29}{c}{\textit{v}\primed} \\
        \cline{3-31}\noalign{\smallskip}
        & &\multicolumn{4}{c}{0} & &\multicolumn{4}{c}{1} & &\multicolumn{4}{c}{2} & &\multicolumn{4}{c}{3} & &\multicolumn{4}{c}{4} & &\multicolumn{4}{c}{5} \\
        \cline{3-31}\noalign{\smallskip}
        \textit{v}\primed\primed & &\multicolumn{2}{c}{a} &\multicolumn{2}{c}{b}  & &\multicolumn{2}{c}{a}  &\multicolumn{2}{c}{b}  & &\multicolumn{2}{c}{a}  &\multicolumn{2}{c}{b}  & &\multicolumn{2}{c}{a}  &\multicolumn{2}{c}{b}  & &\multicolumn{2}{c}{a}  &\multicolumn{2}{c}{b}  & &\multicolumn{2}{c}{a} &\multicolumn{2}{c}{b}\\
        \hline\noalign{\smallskip}
0 &&  3.&047 (-2)  &  3.&069 (-2)  &&  9.&456 (-3)  &  9.&414 (-3)  &&  8.&079 (-4)  &  7.&885 (-4)  &&  1.&094 (-5)  &  9.&656 (-6)  &&  4.&013 (-7)  &  5.&104 (-7)  &&  5.&150 (-8)  &  6.&854 (-8)  \\
1 &&  1.&016 (-2)  &  1.&015 (-2)  &&  1.&350 (-2)  &  1.&374 (-2)  &&  1.&356 (-2)  &  1.&353 (-2)  &&  1.&845 (-3)  &  1.&786 (-3)  &&  2.&298 (-5)  &  1.&838 (-5)  &&  2.&949 (-6)  &  3.&794 (-6)  \\
2 &&  2.&201 (-3)  &  2.&185 (-3)  &&  1.&283 (-2)  &  1.&287 (-2)  &&  4.&951 (-3)  &  5.&146 (-3)  &&  1.&465 (-2)  &  1.&462 (-2)  &&  2.&773 (-3)  &  2.&650 (-3)  &&  2.&247 (-5)  &  1.&407 (-5)  \\
3 &&  3.&937 (-4)  &  3.&878 (-4)  &&  4.&648 (-3)  &  4.&618 (-3)  &&  1.&179 (-2)  &  1.&190 (-2)  &&  1.&263 (-3)  &  1.&379 (-3)  &&  1.&425 (-2)  &  1.&421 (-2)  &&  3.&422 (-3)  &  3.&201 (-3)  \\
4 &&  6.&371 (-5)  &  6.&204 (-5)  &&  1.&171 (-3)  &  1.&154 (-3)  &&  6.&442 (-3)  &  6.&403 (-3)  &&  9.&326 (-3)  &  9.&529 (-3)  &&  1.&068 (-4)  &  1.&417 (-4)  &&  1.&337 (-2)  &  1.&325 (-2)  \\
5 &&  9.&712 (-6)  &  9.&338 (-6)  &&  2.&457 (-4)  &  2.&392 (-4)  &&  2.&145 (-3)  &  2.&113 (-3)  &&  7.&319 (-3)  &  7.&288 (-3)  &&  6.&705 (-3)  &  6.&987 (-3)  &&  4.&296 (-5)  &  2.&853 (-5)  \\ [1ex]
        \hline
        \label{TABfvv}
    \end{tabular}
        \vspace*{-5mm}
    \begin{tablenotes}
        \begin{large}
            \item[a] It should be noted that in the calculation of these values, a wavenumber for the band had to be used. The value chosen was the wavenumber of the P(0) transition. If a different wavenumber is desired, the \Avv\ values in Table \ref{TABAvv} can be used in conjunction with Equation \ref{EQNAvvtofvv} to calculate new \fvv\ values. The numbers in parentheses indicate the exponent.
        \end{large}
    \end{tablenotes}
\end{threeparttable}
}
\end{center}
\end{table}

\begin{figure}
  \hspace*{-2.8cm}
  \includegraphics[width=19cm]{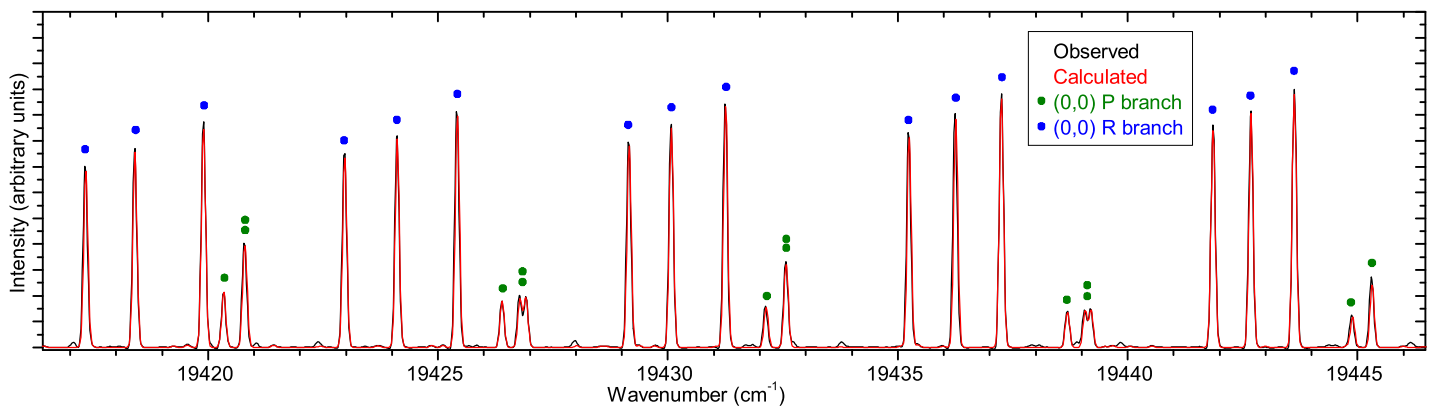}
  \caption{A section of the (0,0) band of the \CC\ Swan system. P branch: \textit{J\primed\primed}=36-41, R branch: \textit{J\primed\primed}=7-13.}\label{fig(0,0)}
\end{figure}

\begin{figure}
  \hspace*{-2.8cm}
  \includegraphics[width=19cm]{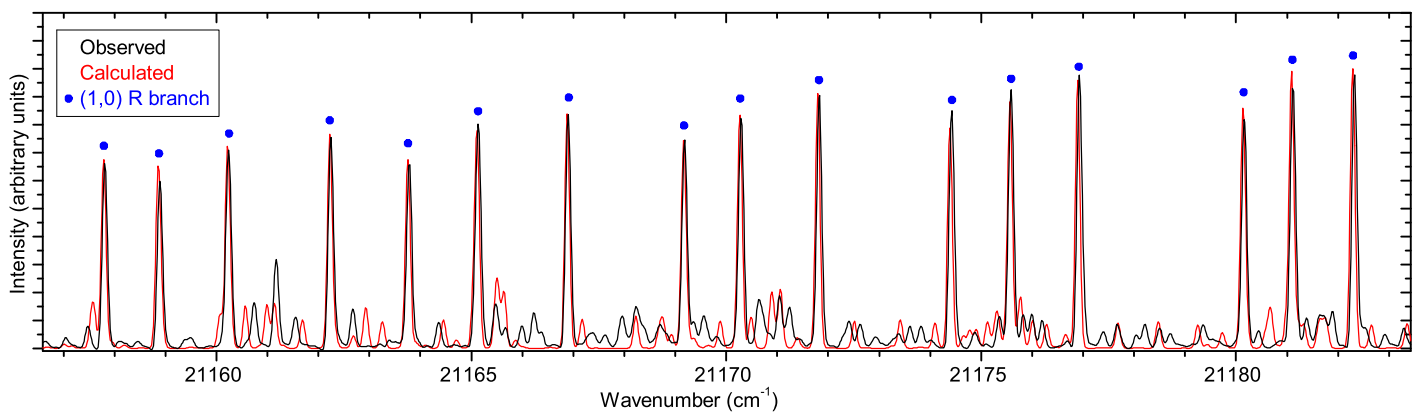}
  \caption{A section of the $\Delta$\emph{v}=+1 sequence of the \CC\ Swan system, showing that the
(1,0) R branch lines match well. The less intense lines do not match as closely. They are a
mixture of the (1,0) P branch, the (9,8) R branch and the (8,7) P branch. (1,0) R branch:
\textit{J\primed\primed}=39-45.}\label{fig(1,0)}
\end{figure}

\begin{figure}
  \hspace*{-2.8cm}
  \includegraphics[width=19cm]{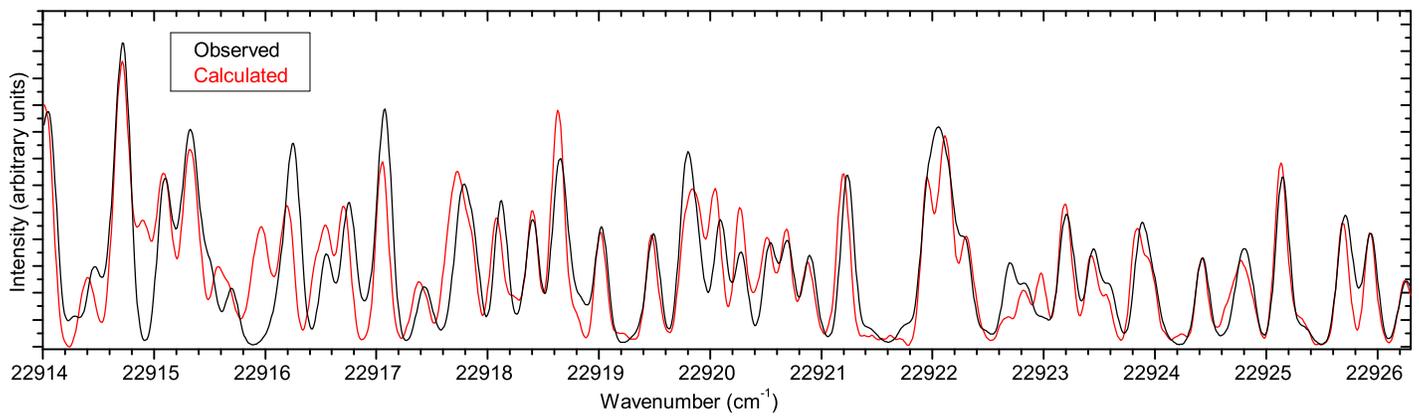}
  \caption{A section of the $\Delta$\emph{v}=+2 sequence of the \CC\ Swan system. Shown is a mixture of the (2,0) R branch, the (3,1) R and P branches, the (4,2) and the P branch,(5,3) P branch,(6,4) P branch.(2,0) R branch: \textit{J\primed\primed}=13-16, (3,1) R branch: \textit{J\primed\primed}=47-51, (3,1) P branch: \textit{J\primed\primed}=2-4, (4,2) P branch: \textit{J\primed\primed}=5-12 and 39-48, (5,3) P branch: \textit{J\primed\primed}=5-17 and 45-52, (6,4) P branch: \textit{J\primed\primed}=1-5}\label{fig(2,0)}
\end{figure}

\end{document}